\documentclass[11pt,a4paper]{article}
\usepackage{amsmath,amssymb}
\usepackage{epsfig,graphicx}

\begin{document}

\title{\bf Analytical properties of the $\pi\pi$ scattering
amplitude and the light scalar mesons}
\author{N.~N.~Achasov$^a$\footnote{{\bf e-mail}: achasov@math.nsc.ru},
\underline{A.~V.~Kiselev}$^{a}$\footnote{{\bf e-mail}:
kiselev@math.nsc.ru}
\\
$^a$ \small{\em Sobolev Institute for Mathematics} \\ \small{\em
Novosibirsk, 630090} }
\date{}
\maketitle

\begin{abstract}
The $\pi\pi$ scattering amplitude with regular analytical
properties in the $s$ complex plane has been constructed. It
describes simultaneously the data on the $\pi\pi$ scattering,
$\phi\to\pi^0\pi^0\gamma$ decay and $\pi\pi\to K\bar K$ reaction.
The chiral shielding of the $\sigma (600)$ meson and it's mixing
with the $f_0(980)$ meson are taken into account also. The data
agrees with the four-quark nature of the $\sigma (600)$ and
$f_0(980)$ mesons.

The amplitude in the range $-5 m_\pi^2 < s < 0.64$ GeV$^2$ also
agrees with results, obtained on the base of the chiral expansion,
dispersion relations and the Roy equations.
\end{abstract}

\section{Introduction}

Study of light scalar resonances is one of the central problems of
non-perturbative QCD, it is important for understanding both the
confinement physics and the chiral symmetry realization way in the
low energy region. The commonly suggested nonet of light scalar
mesons is $f_0(600)$ (or $\sigma (600)$), $K_0^*(800)$ (or $\kappa
(800)$), $f_0(980)$ and $a_0(980)$ Ref. \cite{pdg-2008}.

In the 1987 it was suggested in \cite{achasov-89} to investigate
light scalar mesons in radiative $\phi$ decays. Chiral one-loop
mechanism of the transition $\phi\to K^+K^-\to\gamma f_0(a_0)$
(Kaon Loop model) was also proposed there.

Ten years later, in the 1998, the decays $\phi\to\eta\pi^0\gamma$
and $\phi\to\pi^0\pi^0\gamma$ were experimentally discovered in
Budker INP \cite{snd-cmd}. Then in the 2002 KLOE group published
the high-statistical data on these decays \cite{publ,pi0publ} (800
$\phi\to\eta\pi^0\gamma$ events and 2400 $\phi\to\pi^0\pi^0\gamma$
events).

In \cite{our_f0} the KLOE data on the $\phi\to\pi^0\pi^0\gamma$
decay were described simultaneously with the data on the $\pi\pi$
scattering and the $\pi\pi\to K\bar K$ reaction. The description
was carried out taking into account the chiral shielding
\cite{annshgn-94,annshgn-07} of $\sigma (600)$ and the $\sigma
(600)-f_0(980)$ mixing. The data didn't contradict the existence
of the $\sigma (600)$ meson and yielded evidence in favor of the
four-quark nature of the $\sigma (600)$ and $f_0(980)$ mesons.
These experiments ratified Kaon Loop mechanism .

\begin{figure}
\includegraphics{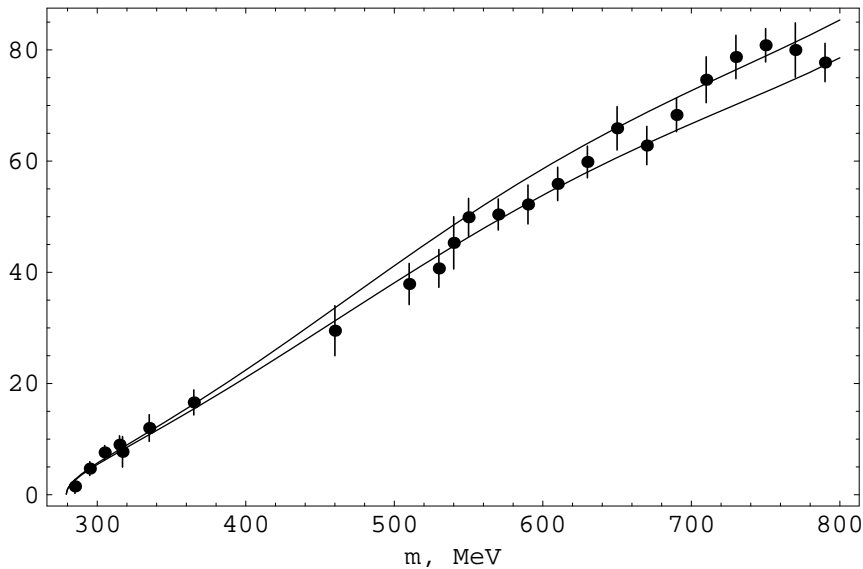}
\caption {The corridor in the phase $\delta_0^0$ of the $\pi\pi$
scattering \cite{sigmaPole} and the experimental data
\cite{pipidata}} \label{fig1}
\end{figure}

\begin{figure}
\includegraphics{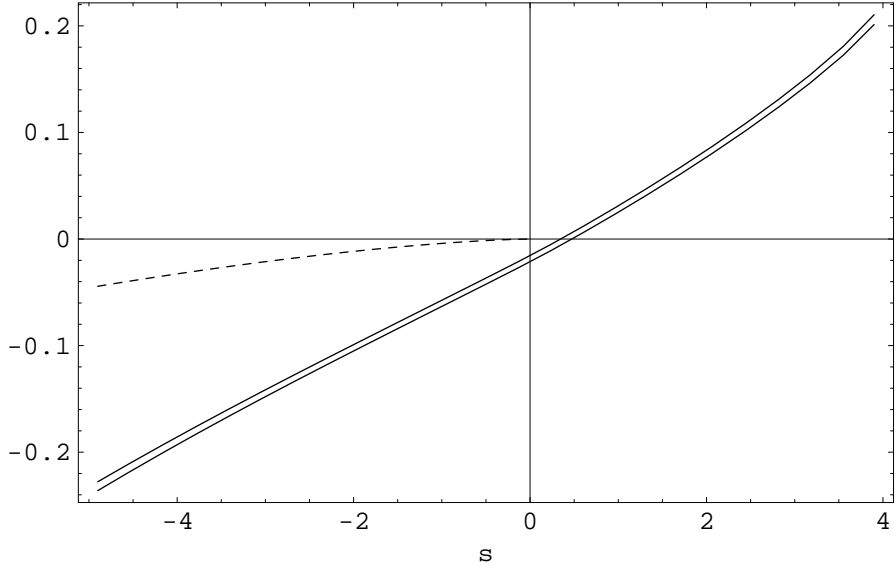}
\caption {The real and the imaginary parts of the amplitude
$T^0_0$ of the $\pi\pi$ scattering ($s$ in units of $m_\pi^2$)
\cite{sigmaPole}} \label{fig2}
\end{figure}

This description revealed new goals. The point is that at the same
time it was calculated in \cite{sigmaPole} the $\pi\pi$ scattering
amplitude in the $s$ complex plane, basing on chiral expansion,
dispersion relations and Roy equations, see Figs.
\ref{fig1},\ref{fig2}. In particular, the pole in $s=M_\sigma^2$
was obtained in \cite{sigmaPole}, where

\begin{equation}
M_\sigma = 441^{+16}_{-8} - i272^{+9}_{-12.5}\ \mbox{MeV}
\label{poleSigma}\end{equation}

\noindent and was assigned to the $\sigma$ resonance.

Aiming the comparison of the results of \cite{our_f0} and
\cite{sigmaPole} it is necessary to build the $\pi\pi$ scattering
amplitude with correct analytical properties in the complex $s$
plane. The point is that in \cite{our_f0} S-matrix of the $\pi\pi$
scattering is the product of the "resonance" and "background"
parts:

$$S_{\pi\pi} = S_{back}\,S_{res}\,, $$

\noindent and the $S_{res}$ had desired analytical properties,
while analytical properties of the $S_{back}$ in the whole complex
$s$ plane were not essential for the aims of \cite{our_f0}, where
mainly physical region was investigated, and Adler zero existence
together with poles absence on the real axis of the $s$ complex
plane were demanded.

In this paper we present the $\pi\pi$ scattering amplitude with
desired analytical properties and the data description obtained
with this amplitude. The comparison with \cite{sigmaPole} results
is presented also.

\section{Some theory}

The S-wave amplitude $T^0_0$ of the $\pi\pi$ scattering with I=0
\cite{achasovEtAl} is

\begin{equation}
T^0_0=\frac{\eta^0_0 e^{2i\delta_0^0}-1}{2i\rho_{\pi\pi}(m)}=
\frac{e^{2i\delta_B^{\pi\pi}}-1}{2i\rho_{\pi\pi}(m)}+
e^{2i\delta_B^{\pi\pi}}\sum_{R,R'}\frac{g_{R\pi\pi}G_{RR'}^{-1}g_{R'\pi\pi}}{16\pi}\,.
\label{pipiamp}\end{equation}

Here $\eta^0_0\equiv \eta^0_0(m)$ is the inelasticity,
$\delta_0^0$ is the scattering phase, $\delta_B^{\pi\pi}$ is the
phase of the elastic background, and

\[G_{RR'}(m)=\left( \begin{array}{cc} D_{f_0}(m)&-\Pi_{f_0\sigma}(m)\\-\Pi_{f_0
\sigma}(m)&D_{\sigma}(m)\end{array}\right)\]

The desired analytical properties of the $\pi\pi$ scattering
amplitude are: two cuts in the $s$-complex plane, Adler zero,
absence of poles on the physical sheet of the Riemannian surface,
resonance poles on the second sheet of the Riemannian surface.
This applies curtain restrictions on the $\delta_B^{\pi\pi}$.

The inverse propagator of scalar $R$ \cite{adsh-79}

\begin{equation}
 D_R(m^2) = m_R^2-m^2+ Re\left
(\Pi_R(m_R^2)\right )-\Pi_R(m^2) \label{propagator}\end{equation}

\noindent where

$$\Pi_R^{ab}(m^2)=\frac{1}{\pi}[m^2-(m_a+m_b)^2]\int_{(m_a+m_b)^2}^{\infty}\frac{\bar{m}\Gamma(R\to
ab,\bar{m})\,
d\bar{m}^2}{[\bar{m}^2-(m_a+m_b)^2](\bar{m}^2-m^2-i\varepsilon)}$$

So, following \cite{our_prop}, we have

$$ Im\left (D_R(z)\right )= -y\left (1
+\sum_{ab}\frac{1}{\pi}\int_{(m_a+m_b)^2}^{\infty}\frac{\bar{m}\Gamma(R\to
ab, \bar{m})}{|\bar{m}^2- z|^2}\, d\bar{m}^2\right )$$

\noindent and, since expression in brackets is positive,
Im$D_R(z)$ may be equal to zero only when $y\equiv$Im~$z=0$.

Let's use the same mechanism and take $e^{2i\delta_B^{\pi\pi}}$ in
the form

$$e^{2i\delta_B^{\pi\pi}}=\frac{P^*_{\pi 1}(s)P^*_{\pi
2}(s)}{P_{\pi 1}(s)P_{\pi 2}(s)}\,,$$

\noindent where

$$P_{\pi 1}(s) =
a_1-a_2\frac{s}{4m_\pi^2}-\Pi_{\pi\pi}(s)+a_3\,\Pi_{\pi\pi}(4m_\pi^2-s)-a_4
Q_1(s)\,,$$

$$Q_1(s)
=\frac{1}{\pi}\int_{4m_\pi^2}^{\infty}\frac{s-4m_\pi^2}{s'-4m_\pi^2}\,\frac{\rho_{\pi\pi}(s')}{s'-s-i\varepsilon}K_1(s')$$

$$ Im(Q_1(s)) = K_1(s)\rho_{\pi\pi}(s)$$\\[1mm]

Since Im$(Q_1(s)) = K_1(s)\rho_{\pi\pi}(s)$ we require that
$K_1(s)$ should be positive and have no singularities on the
physical sheet. This is provided if we take

$$K_1(s) =
\frac{L_1(s)}{D_1(4m_\pi^2-s)D_2(4m_\pi^2-s)D_3(4m_\pi^2-s)D_4(4m_\pi^2-s)}\,,$$

$$D_i(s)=m_{i}^2-s-g_i\Pi_{\pi\pi}(s)\,,$$

$$L_1(s) = (s-4m_\pi^2)^4+\alpha_1 4m_\pi^2
(s-4m_\pi^2)^3+\alpha_2 (4m_\pi^2)^2 (s-4m_\pi^2)^2+$$

$$+\alpha_3 (4m_\pi^2)^3
(s-4m_\pi^2)+q_1^8+\sqrt{s}\bigg(c_1(2m_\pi)^7+c_2(2m_\pi)^5(s-4m_\pi^2)+$$

$$+c_3(2m_\pi)^3(s-4m_\pi^2)^2+c_4(2m_\pi)(s-4m_\pi^2)^3\bigg)$$

$$ P_{\pi 2}(s) = \frac{\Lambda^2+s-4m_\pi^2}{4m_\pi^2} +
k_2Q_2(s)\,, $$

$$Q_2(s)
=\frac{1}{\pi}\int_{4m_\pi^2}^{\infty}\frac{s-4m_\pi^2}{s'-4m_\pi^2}\,\frac{\rho_{\pi\pi}(s')}{s'-s-i\varepsilon}K_2(s')\,,
$$

$$K_2(s) =
\frac{L_2(s)}{D_{1A}(4m_\pi^2-s)D_{2A}(4m_\pi^2-s)D_{3A}(4m_\pi^2-s)}\,,$$

$$L_2(s) = 4m_\pi^2\bigg(s^2+\beta (4m_\pi^2) s + \gamma_1
(2m_\pi)^3 s^{1/2} + \gamma_2 (2m_\pi) s^{3/2}\bigg)$$

\begin{figure}
\includegraphics[width=110 mm, height=66 mm]{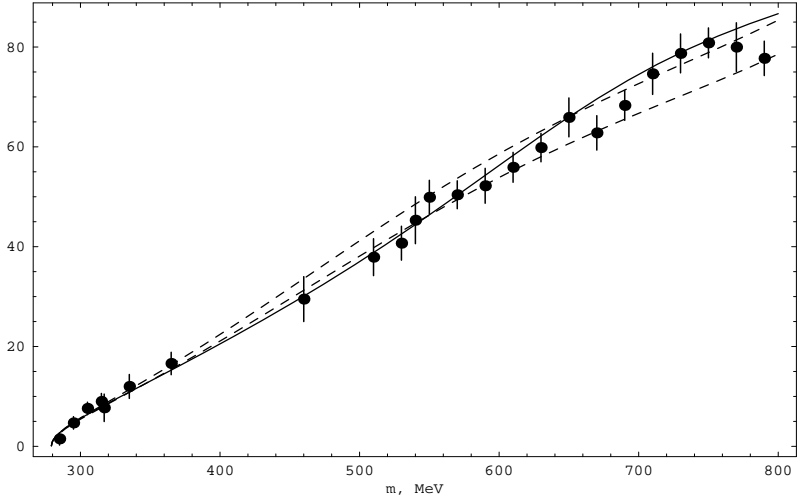}
\caption {The phase $\delta_0^0$ of the $\pi\pi$ scattering. Solid
line is our description, dashed lines mark borders of the corridor
\cite{sigmaPole}, points are experimental data} \label{fig3}
\end{figure}

\begin{figure}
\includegraphics[width=110 mm, height=66 mm]{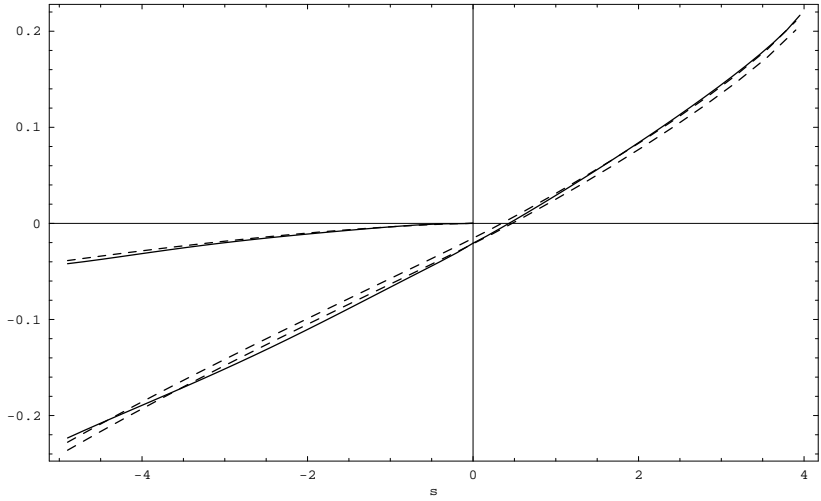}
\caption {The real and the imaginary parts of the amplitude
$T^0_0$ of the $\pi\pi$ scattering ($s$ in units of $m_\pi^2$).
Solid lines show our description, dashed lines mark borders of the
real part corridor and the imaginary part for $s < 0$
\cite{sigmaPole}} \label{fig4}
\end{figure}

\begin{figure}
\includegraphics{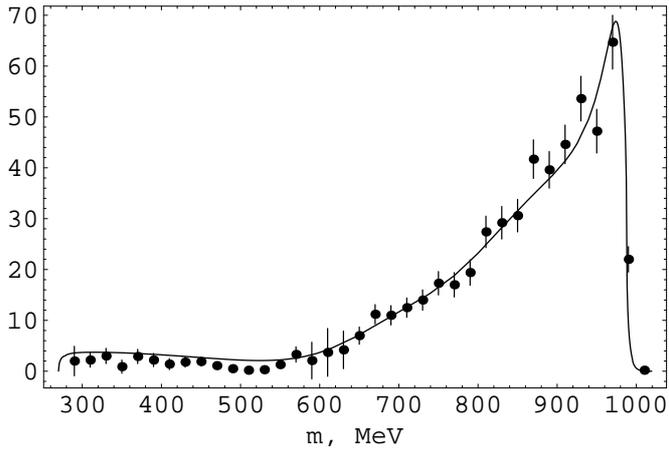}
\caption {The $\pi^0\pi^0$ spectrum in $\phi\to\pi^0\pi^0\gamma$
decay, solid line is our description} \label{fig5}
\end{figure}

With the help of the above formulas we obtain the results, shown
in Figs \ref{fig3},\ref{fig4},\ref{fig5}. One can see that the
experimental data and the corridor from \cite{sigmaPole} are
described well.

\begin{figure}
\includegraphics{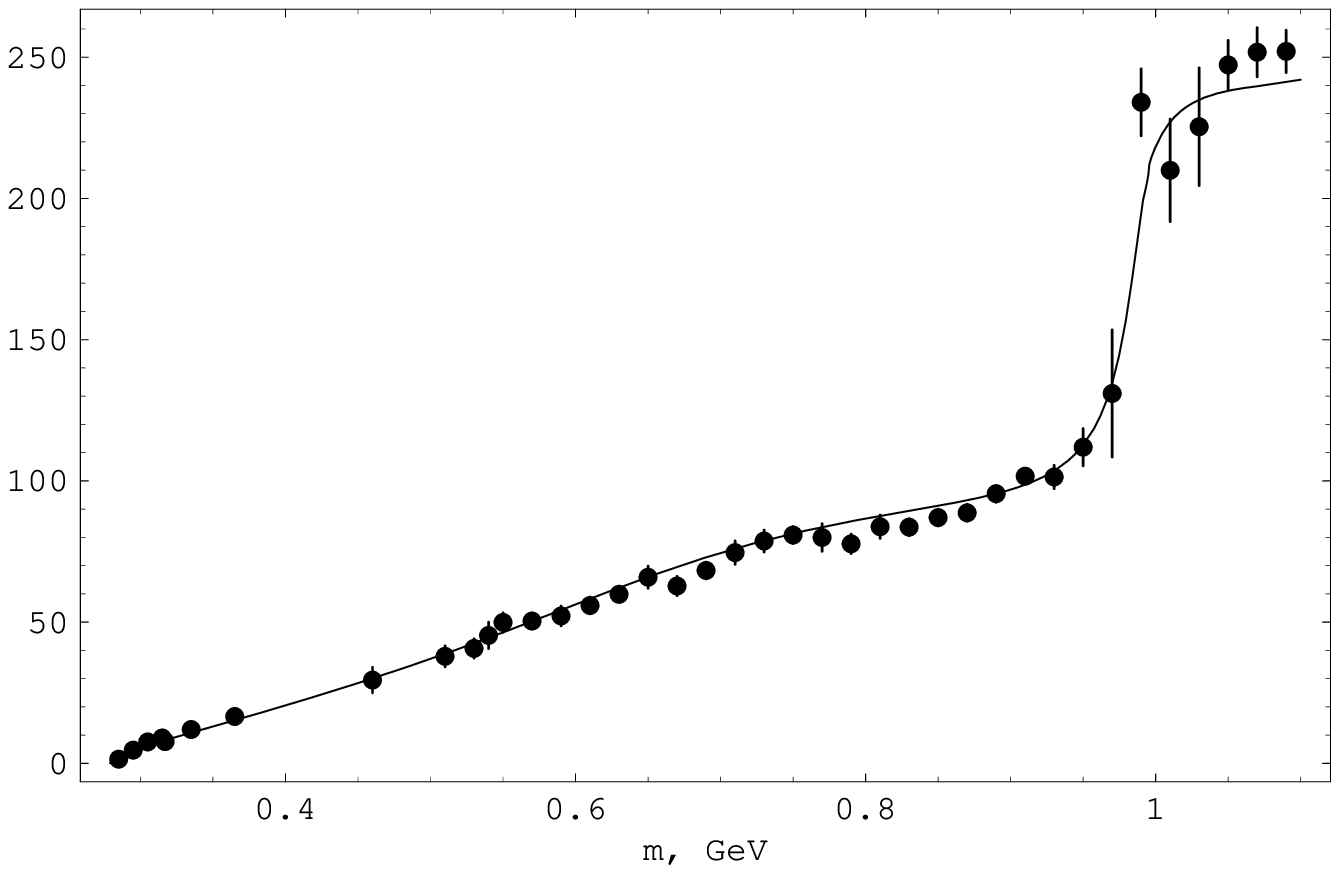} \caption {The phase $\delta_0^0$ of the $\pi\pi$
scattering, solid line is our description} \label{fig6}
\end{figure}

\section{Poles of the resonances}

The $\sigma$ pole is not at (\ref{poleSigma}) but at $$M_\sigma =
617 - i318\ \mbox{MeV}$$

This deviation is not a crime, because Roy equations are
approximate, they are one-channel ($\pi\pi$) \footnote{Note that
for one-channel $\sigma$ inverse propagator
$$D_\sigma(m)=m^2_\sigma - m^2 + Re
\big(\Pi_{\sigma}^{\pi\pi}(m^2_\sigma)\big)-\Pi_{\sigma}^{\pi\pi}(m^2)$$

\noindent the pole at (\ref{poleSigma}) means violation of the
K\"allen -- Lehmann representation, see \cite{our_prop}.}.

Also $M_{f_0} = 1188 - i780\ \mbox{MeV}$, while $m_{f_0} = 986\
\mbox{MeV}, \Gamma(f_0\to\pi\pi,m_{f_0}) = 85\ \mbox{MeV}$.

We find out that the role of high channels is very important for
pole positions. For R=$\sigma(600),\ f_0(980)$ we take
$$\Pi_R=\Pi_R^{\pi^+\pi^-}+\Pi_R^{\pi^0\pi^0}+\Pi_R^{K^+K^-}+
\Pi_R^{K^0\bar{K^0}}+\Pi_R^{\eta \eta}+\Pi_R^{\eta '
\eta}+\Pi_R^{\eta ' \eta '}$$

But a fit without $f_0$ coupling to $\eta \eta, \eta \eta'$, and
$\eta' \eta'$ and small $\sigma$ coupling to these channels gives
us

\noindent $M_\sigma = 639 - i313\ \mbox{MeV}$ (instead of $617 -
i318\ \mbox{MeV}$),

\noindent $M_{f_0} = 984 - i423\ \mbox{MeV}$ (instead of $1188 -
i780\ \mbox{MeV}$).\\[1mm]

Note that if we additionally neglect $\sigma$ coupling to the
$K\bar K$ in the obtained results, the pole is transferred to
$M_\sigma = 562 - i233\ \mbox{MeV}$.

\section{Perspectives}

We plan to use results of this work for study of other reactions,
including decays of heavy quarkoniums and pion polarization.

Special words should be said about refinement of the obtained
results. New precise measurement of the inelasticity $\eta^0_0$
near $K\bar K$ threshold would be crucial, it can improve our
knowledge about light scalar mesons a lot. In fact, commonly used
data on $\eta^0_0$ were obtained forty years ago, and they are not
enough. As far as we know, some collaborations even have raw data
(for example, VES Collaboration), and all they have to do is to
process these data.

New precise data on the $\phi\to\pi^0\pi^0\gamma$ decay would also
clarify the situation. The KLOE Collaboration has measured this
reaction with excellent precision, and hopefully they will publish
final results.

\section{Summary}

\begin{enumerate}

\item The $\pi\pi$ scattering amplitude with correct analytical properties
has been built.

\item This amplitude describes experimental data and the results based on Roy equations on the real $s$
axis.

\item A contradiction in position of $\sigma$ pole may be an
indication of the important role of high channels in the $\pi\pi$
scattering amplitude and its analytical continuation even for
$|s|$ much less than $m_{f_0}$.

\item The $f_0$ pole is situated far from the position, predicted by
Breit-Wigner approximation.

\item New experiments are important for the study of light scalar
mesons.

\end{enumerate}

\section{Acknowledgements}
We are grateful to H. Leutwyler for useful intensive
communications, in particular for tables with $\pi\pi$ scattering
amplitude on the real $s$ axis, obtained in \cite{sigmaPole}. This
work was supported in part by RFBR, Grant No 10-02-00016. A.V.K.
was also supported by RFBR Grant No 10-02-16010-mob\_s\_ros.

\end{document}